\documentclass[amsmath,amssymb,twocolumn,prc,showpacs,nofootinbib,floatfix,superscriptaddress]{revtex4-1}

\usepackage{breakcites}
\usepackage{float}
\usepackage{soul}
\usepackage{amsmath}
\usepackage{graphicx}
\usepackage{dcolumn}
\usepackage{bm}
\usepackage{natbib}
\usepackage{tikz}
\usepackage{color}
\usepackage{multirow}
\usepackage{subfigure}
\usepackage{longtable}
\usepackage{dcolumn}

\usepackage[colorlinks, linkcolor={blue},citecolor={blue},breaklinks]{hyperref}

\begin{document}

\preprint{APS/123-QED}
\title{First application of Markov Chain Monte Carlo-based Bayesian data analysis to the Doppler-Shift Attenuation Method}
\thanks{These authors contributed equally to this work and should be considered co-first authors.\label{Coauthors}}%

\author{L.~J.~Sun\textsuperscript{\ref{Coauthors}}}
\email{sunli@frib.msu.edu}
\affiliation{National Superconducting Cyclotron Laboratory, Michigan State University, East Lansing, Michigan 48824, USA}
\affiliation{School of Physics and Astronomy, Shanghai Jiao Tong University, Shanghai 200240, China}
\author{C.~Fry\textsuperscript{\ref{Coauthors}}}
\email{cfry@lanl.gov}
\affiliation{National Superconducting Cyclotron Laboratory, Michigan State University, East Lansing, Michigan 48824, USA}
\affiliation{Department of Physics and Astronomy, Michigan State University, East Lansing, Michigan 48824, USA}
\affiliation{Los Alamos National Laboratory, Los Alamos, New Mexico 87545, USA}
\author{B.~Davids}
\email{davids@triumf.ca}
\affiliation{TRIUMF, 4004 Wesbrook Mall, Vancouver, British Columbia V6T 2A3, Canada}
\affiliation{Department of Physics, Simon Fraser University, Burnaby, British Columbia V5A 1S6, Canada}
\author{N.~Esker}
\affiliation{TRIUMF, 4004 Wesbrook Mall, Vancouver, British Columbia V6T 2A3, Canada}
\affiliation{Department of Chemistry, San Jos\'{e} State University, San Jose, California 95192, USA}
\author{C.~Wrede}
\email{wrede@nscl.msu.edu}
\affiliation{Department of Physics and Astronomy, Michigan State University, East Lansing, Michigan 48824, USA}
\affiliation{National Superconducting Cyclotron Laboratory, Michigan State University, East Lansing, Michigan 48824, USA}
\author{M.~Alcorta}
\affiliation{TRIUMF, 4004 Wesbrook Mall, Vancouver, British Columbia V6T 2A3, Canada}
\author{S.~Bhattacharjee}
\affiliation{TRIUMF, 4004 Wesbrook Mall, Vancouver, British Columbia V6T 2A3, Canada}
\author{M.~Bowry}
\affiliation{TRIUMF, 4004 Wesbrook Mall, Vancouver, British Columbia V6T 2A3, Canada}
\author{B.~A.~Brown}
\affiliation{National Superconducting Cyclotron Laboratory, Michigan State University, East Lansing, Michigan 48824, USA}
\affiliation{Department of Physics and Astronomy, Michigan State University, East Lansing, Michigan 48824, USA}
\author{T.~Budner}
\affiliation{National Superconducting Cyclotron Laboratory, Michigan State University, East Lansing, Michigan 48824, USA}
\affiliation{Department of Physics and Astronomy, Michigan State University, East Lansing, Michigan 48824, USA}
\author{R.~Caballero-Folch}
\affiliation{TRIUMF, 4004 Wesbrook Mall, Vancouver, British Columbia V6T 2A3, Canada}
\author{L.~Evitts}
\affiliation{TRIUMF, 4004 Wesbrook Mall, Vancouver, British Columbia V6T 2A3, Canada}
\author{M.~Friedman}
\affiliation{National Superconducting Cyclotron Laboratory, Michigan State University, East Lansing, Michigan 48824, USA}
\affiliation{Racah Institute of Physics, Hebrew University, Jerusalem 91904, Israel}
\author{A.~B.~Garnsworthy}
\affiliation{TRIUMF, 4004 Wesbrook Mall, Vancouver, British Columbia V6T 2A3, Canada}
\author{B.~E.~Glassman}
\affiliation{National Superconducting Cyclotron Laboratory, Michigan State University, East Lansing, Michigan 48824, USA}
\affiliation{Department of Physics and Astronomy, Michigan State University, East Lansing, Michigan 48824, USA}
\author{G.~Hackman}
\affiliation{TRIUMF, 4004 Wesbrook Mall, Vancouver, British Columbia V6T 2A3, Canada}
\author{J.~Henderson}
\affiliation{TRIUMF, 4004 Wesbrook Mall, Vancouver, British Columbia V6T 2A3, Canada}
\author{O.~S.~Kirsebom}
\affiliation{Department of Physics and Atmospheric Science, Dalhousie University, Halifax, Nova Scotia B3H 4R2, Canada}
\author{A.~Kurkjian}
\affiliation{TRIUMF, 4004 Wesbrook Mall, Vancouver, British Columbia V6T 2A3, Canada}
\author{J.~Lighthall}
\affiliation{TRIUMF, 4004 Wesbrook Mall, Vancouver, British Columbia V6T 2A3, Canada}
\author{P.~Machule}
\affiliation{TRIUMF, 4004 Wesbrook Mall, Vancouver, British Columbia V6T 2A3, Canada}
\author{J.~Measures}
\affiliation{TRIUMF, 4004 Wesbrook Mall, Vancouver, British Columbia V6T 2A3, Canada}
\author{M.~Moukaddam}
\affiliation{TRIUMF, 4004 Wesbrook Mall, Vancouver, British Columbia V6T 2A3, Canada}
\author{J.~Park}
\affiliation{TRIUMF, 4004 Wesbrook Mall, Vancouver, British Columbia V6T 2A3, Canada}
\affiliation{Center for Exotic Nuclear Studies, Institute for Basic Science, Daejeon 34126, Republic of Korea}
\author{C.~Pearson}
\affiliation{TRIUMF, 4004 Wesbrook Mall, Vancouver, British Columbia V6T 2A3, Canada}
\author{D.~P\'erez-Loureiro}
\affiliation{Canadian Nuclear Laboratories, Chalk River, Ontario K0J 1J0, Canada}
\author{C.~Ruiz}
\affiliation{TRIUMF, 4004 Wesbrook Mall, Vancouver, British Columbia V6T 2A3, Canada}
\author{P.~Ruotsalainen}
\affiliation{TRIUMF, 4004 Wesbrook Mall, Vancouver, British Columbia V6T 2A3, Canada}
\author{J.~Smallcombe}
\affiliation{TRIUMF, 4004 Wesbrook Mall, Vancouver, British Columbia V6T 2A3, Canada}
\author{J.~K.~Smith}
\affiliation{TRIUMF, 4004 Wesbrook Mall, Vancouver, British Columbia V6T 2A3, Canada}
\author{D.~Southall}
\affiliation{TRIUMF, 4004 Wesbrook Mall, Vancouver, British Columbia V6T 2A3, Canada}
\author{J.~Surbrook}
\affiliation{National Superconducting Cyclotron Laboratory, Michigan State University, East Lansing, Michigan 48824, USA}
\affiliation{Department of Physics and Astronomy, Michigan State University, East Lansing, Michigan 48824, USA}
\author{M.~Williams}
\affiliation{TRIUMF, 4004 Wesbrook Mall, Vancouver, British Columbia V6T 2A3, Canada}
\affiliation{Department of Physics, University of York, Heslington, York YO10 5DD, United Kingdom}
\author{L.~E.~Weghorn}
\affiliation{National Superconducting Cyclotron Laboratory, Michigan State University, East Lansing, Michigan 48824, USA}
\affiliation{Department of Physics and Astronomy, Michigan State University, East Lansing, Michigan 48824, USA}
\date{\today}
\begin{abstract}
Motivated primarily by the large uncertainties in the thermonuclear rate of the $^{30}$P$(p,\gamma)^{31}$S reaction that limit our understanding of classical novae, we carried out lifetime measurements of $^{31}$S excited states using the Doppler Shift Lifetimes (DSL) facility at the TRIUMF Isotope Separator and Accelerator (ISAC-II) facility. The $^{31}$S excited states were populated by the $^{3}$He$(^{32}$S$,\alpha)^{31}$S reaction. The deexcitation $\gamma$ rays were detected by a clover-type high-purity germanium detector in coincidence with the $\alpha$ particles detected by a silicon detector telescope. We have applied modern Markov chain Monte Carlo-based Bayesian methods to perform lineshape analyses of Doppler-shift attenuation method $\gamma$-ray data for the first time. We have determined the lifetimes of the two lowest-lying $^{31}$S excited states. First experimental upper limits on the lifetimes of four higher-lying states have been obtained. The experimental results were compared to shell-model calculations using five universal $sd$-shell Hamiltonians. Evidence for $\gamma$ rays originating from the astrophysically important $J^\pi=3/2^+$, 260-keV $^{30}$P$(p,\gamma)^{31}$S resonance has also been observed, although strong constraints on the lifetime will require better statistics.
\end{abstract}
\maketitle

\section{Introduction}
Classical novae are one of the most frequent thermonuclear stellar explosions in the Galaxy. They are powered by thermonuclear runaways occurring in the accreted envelope transferred from a companion star onto a compact white dwarf in a close binary system~\cite{Jose_2016,Iliadis_2015}. In classical novae, the $^{30}$P$(p,\gamma)^{31}$S reaction acts as a nucleosynthesis bottleneck in the flow of material to heavier masses~\cite{Jose_NPA2006}. The large uncertainty in the $^{30}$P$(p,\gamma)^{31}$S rate impacts the identification of certain presolar nova grains~\cite{Jose_APJ2004}, the calibration of nuclear nova thermometers~\cite{Downen_APJ2013}, and the Si/H abundance ratio, which can be used to constrain the degree of mixing between the white dwarf's outer layers and the accreted envelope~\cite{Kelly_APJ2013}. It is not currently possible to measure the $^{30}$P$(p,\gamma)^{31}$S reaction directly because intense low energy $^{30}$P beams are not available. The thermonuclear rate of the $^{30}$P$(p,\gamma)^{31}$S reaction over most of the peak nova temperatures (0.1$-$0.4~GK) is found to be dominated by proton capture into a 260-keV $3/2^+$ resonance with an excitation energy of $E_x=6390.2(7)$~keV in $^{31}$S~\cite{Wrede_AIPA2014,Bennett_PRL2016}. Recent experimental work has unambiguously determined the energy, the spin and parity, and the proton-decay branching ratio of this resonance~\cite{Bennett_PRL2016,Bennett_PRC2018,Budner_PRL2022}, leaving the lifetime as the final missing piece of the puzzle. So far, the lifetimes of three relatively long-lived $^{31}$S states at 1248~\cite{Engmann_NPA1971,Doornenbal_NIMA2010,Herlitzius_Thesis2013,Tonev_PLB2021}, 2234~\cite{Engmann_NPA1971}, and 4451~keV~\cite{Tonev_PLB2021,Pattabiraman_PRC2008} have been reported. The main scientific goal of this work is to expand lifetime measurements to more excited states in $^{31}$S, including the $3/2^+$ state at 6390~keV using the Doppler Shift Attenuation Method ($\mbox{DSAM}$).

Lifetime measurements using $\gamma$-ray spectroscopy provide not only important input for astrophysical models but also a sensitive benchmark for nuclear structure models. $\mbox{DSAM}$ is a widely-used method for measuring lifetimes of excited nuclear states in the fs to ps range~\cite{Elliott_PR1948,Devons_PPSA1955,Devons_PPSA1956}. {\color{black}Despite the wide use of this method, a unified treatment of all the uncertainties associated with systematic effects has been a long-standing issue. Note that the classical frequentist approach ($\chi^2$ minimization) does not in itself provide any uncertainty, but it is a common practice to assume a normal distribution and vary each parameter by one standard deviation while fixing other parameters at some plausible values. Uncertainties from different sources are often assessed independently and then added in quadrature to obtain the total uncertainty. The statistical meaning is even less rigorous when combining upper/lower limits instead of finite values. Multiple parameters often have complex interrelationships, and their correlations may be underestimated or overestimated by standard frequentist approaches~\cite{King_PRL2019,Catacora-Rios_PRC2021,Lovell_JPG2021}.

Loosely speaking, an inverse problem is where we observe an effect and want to determine the cause~\cite{Colin_2013}. Inferring lifetimes from observed $\gamma$-ray spectra is such an inverse problem and represents an ideal case for the application of Bayes's theorem. Bayesian statistics offers natural parameter estimation methods with faithful assessments of uncertainty~\cite{Bayes_PT1763,Jeffreys_1939}. Owing to the distribution complexity and high dimensionality, practical use of Bayesian statistics often requires Markov chain Monte Carlo (\mbox{MCMC})~\cite{Metropolis_JCP1953,Hastings_Bio1970}, an efficient sampling method to systematically explore complex high-dimensional parameter spaces~\cite{Sharma_ARAA2017,Schoot_NRMP2021}. With the advent of modern computational power, there has been a surge of interest in incorporating Bayesian and MCMC techniques in nuclear physics, in particular, the studies of heavy-ion collisions~\cite{Novak_PRC2014,Pratt_PRL2015,Bernhard_PRC2016,Bernhard_NP2019,Morfouace_PLB2019,Moreland_PRC2020,Xie_JPG2021} and low-energy nuclear reactions~\cite{King_PRL2019,Catacora-Rios_PRC2021,Lovell_JPG2021,Lovell_PRC2018,Yang_PLB2020,Marshall_PRC2020,Pruitt_PRC2020,Iliadis_APJ2016,Zhang_Nature2022}. Although Bayesian methods are playing increasingly important roles in many aspects of nuclear physics~\cite{Phillips_JPG2021,Bedaque_EPJA2021,Boehnlein_RMP2022}, to the best of our knowledge, no one had performed $\mbox{DSAM}$ lifetime data analysis within a Bayesian framework. Our previous work~\cite{Galinski_PRC2014} took the very first step in that direction. In this Letter, we further apply \mbox{MCMC}-based Bayesian parameter estimation methods to $\mbox{DSAM}$ lineshape analyses, providing a reliable uncertainty quantification in a multi-dimensional parameter space.}

\section{Experiment}
The experiment was done using the Doppler Shift Lifetimes (DSL) chamber~\cite{Davids_HI2014} specifically designed for $\mbox{DSAM}$ experiments~\cite{Galinski_PRC2014,Kanungo_PRC2006,Mythili_PRC2008,Kirsebom_PRC2016} at the ISAC-II facility of TRIUMF. A 128-MeV $^{32}$S$^{7+}$ beam bombarded a $^3$He-implanted Au target and the excited states in $^{31}$S were populated via the $^{3}$He$(^{32}$S$,\alpha)^{31}$S reaction. We employed inverse kinematics to ensure a large Doppler shift in the $\gamma$-ray spectra. The $\alpha$ particles were detected using a silicon detector telescope placed downstream of the target. The telescope consisted of two ORTEC B Series Si surface barrier detectors with an active area of 150~mm$^2$ and thicknesses of 87~$\mathrm{\mu}$m and 1~mm, respectively~\cite{ORTEC_B}. An aperture was placed in front of the telescope, limiting the ejectile acceptance angle to $<$$13^{\circ}$. Deexcitation $\gamma$ rays were detected in coincidence with $\alpha$-particles by using a clover-type high-purity germanium detector~\cite{Rizwan_NIMA2016,Garnsworthy_NIMA2019} at a distance of 78~mm from the target, centered at 0$^{\circ}$ with respect to the beam axis. See Supplemental Material for more technical details.

The Si detectors were calibrated using a source containing $^{239}$Pu, $^{241}$Am, and $^{244}$Cm, with strong $\alpha$ lines at 5.155~MeV, 5.486~MeV, and 5.805~MeV. A linear calibration was applied and used to extrapolate to higher energies. The extrapolation was verified by comparing the energy loss of punch-through particles to \textsc{srim} calculations~\cite{Ziegler_NIMB2010}. A $^{56}$Co source was used initially to calibrate the Ge detector. A line from $^{197}$Au Coulomb excitation at 279.01(5)~keV~\cite{Huang_NDS2005} and a line from $^{39}$K produced in $^{32}$S+$^{12}$C fusion evaporation at 2814.06(20)~keV~\cite{Chen_NDS2018} were observed with high statistics. The vast majority of the $\gamma$ rays constituting these lines were emitted after the recoils stopped; hence, they are unshifted and used as run-by-run calibration standards. The accuracy of the calibration at high energies was verified by a 6128.63(4)-keV $\gamma$ ray originating from the deexcitation of the second excited state in $^{16}$O~\cite{Tilley_NPA1993}. The energies deposited in all four crystals of the clover detector were summed together to increase the photo-peak efficiency while reducing the Compton scattering background~\cite{Rizwan_NIMA2016}. Lifetimes of $^{31}$S states were then determined from a lineshape analysis of this addback spectrum.

\begin{figure}
\begin{center}
\includegraphics[width=8.6cm]{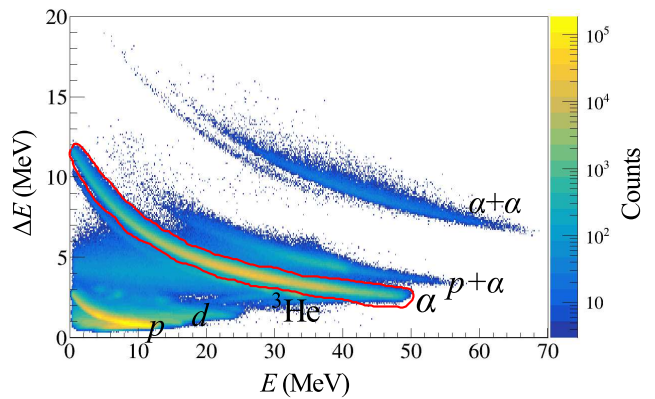}
\caption{\label{PID}Particle identification plot of the energy loss ($\Delta E$) in the 87-$\mathrm{\mu}$m Si detector versus the residual energy deposited in the 1-mm Si detector ($E$). Each locus of points represents a charged particle group or a coincidence summing of two groups. The red contour encloses the $\alpha$ particles of interest.}
\end{center}
\end{figure}

The Si detector telescope particle identification plot is shown in Fig.~\ref{PID}. The $\alpha$-particle group is separated from other charged particle groups. By gating on $\alpha$ particles with specific energies calculated by relativistic reaction kinematics, we suppressed competing reaction channels and indirect feedings from higher-lying levels to ensure a direct population by the transfer reaction, resulting in significantly cleaner $\gamma$-ray spectra.

\section{Simulation \& Results}
The $\gamma$-ray lineshape is sensitive to the $^{31}$S velocity distribution and all other physical effects, and therefore a lineshape analysis is more rigorous and gives more information than a centroid-shift analysis. Detailed Monte Carlo simulations were written to model Doppler-shifted lineshapes for fs lifetimes~\cite{Galinski_PRC2014,Kirsebom_PRC2016}. A new Monte Carlo simulation using \textsc{geant}4~\cite{Agostinelli_NIMA2003,Allison_NIMA2016} was developed in this work to model lineshapes for ps lifetimes as well. We began by sampling the position where the transfer reaction happens from a uniform circular transverse profile defined by the beam spot and the $^{3}$He implantation depth profile calculated by \textsc{srim}~\cite{Ziegler_NIMB2010}. The kinetic energy of the beam was sampled from a Gaussian beam energy distribution with a spread of 0.2\% (full width at half maximum) and energy loss in the target based on the reaction location. The emission angle of the $\alpha$ particle was chosen randomly from an isotropic distribution in the laboratory frame. The error introduced by this simplifying assumption was estimated by trying a few different realistic anisotropic distributions and was found to be rather small due to the limited angular acceptance~\cite{Galinski_PRC2014,Kirsebom_PRC2016}. The energy and momentum of the emitted $\alpha$ particle were calculated using relativistic kinematics from the $Q$-value of the transfer reaction and the kinetic energy of the beam. The $Q$-value of the transfer reaction depends on the populated state in $^{31}$S as $Q=Q_0-E_\mathrm{ex}$ with $Q_0=5.533$~MeV corresponding to the ground state and $E_\mathrm{ex}$ the excitation energy of the populated state in $^{31}$S~\cite{Wang_CPC2021}. We then determine the 4-momentum of the excited $^{31}$S recoil. If a $\gamma$ ray is emitted while the $^{31}$S recoil is still moving, it will be Doppler shifted in the laboratory frame. A detector response of the form of an exponentially modified Gaussian (EMG) function~\cite{Glassman_PRC2019,Sun_PRC2021} was added to the $\gamma$-ray energy recorded by the germanium at the end. The decay and width parameters of the EMG function were empirically characterized as a function of energy by fitting unshifted $\gamma$-ray peaks originating from long-lived states populated by Coulomb excitation and fusion-evaporation reactions at energies of 279.01(5) and 547.5(3)~keV [$^{197}$Au]~\cite{Huang_NDS2005}, 2814.06(20) and 3597.26(25)~keV [$^{39}$K]~\cite{Chen_NDS2018}, 3736.5(3)~keV [$^{40}$Ca]~\cite{Chen_NDS2017}, and 6128.63(4)~keV [$^{16}$O]~\cite{Tilley_NPA1993}.

\begin{figure}
\begin{center}
\includegraphics[width=8.6cm]{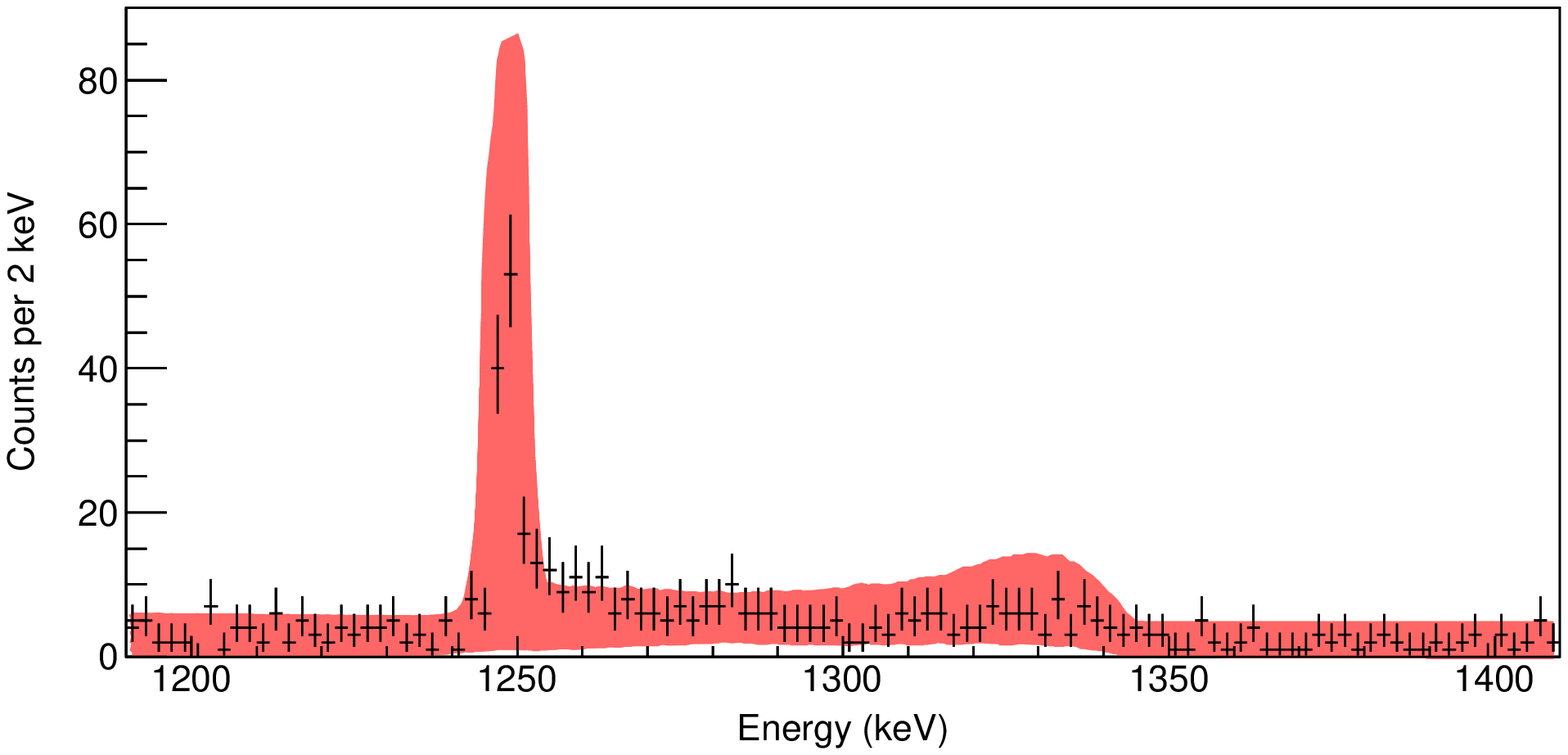}
\includegraphics[width=8.6cm]{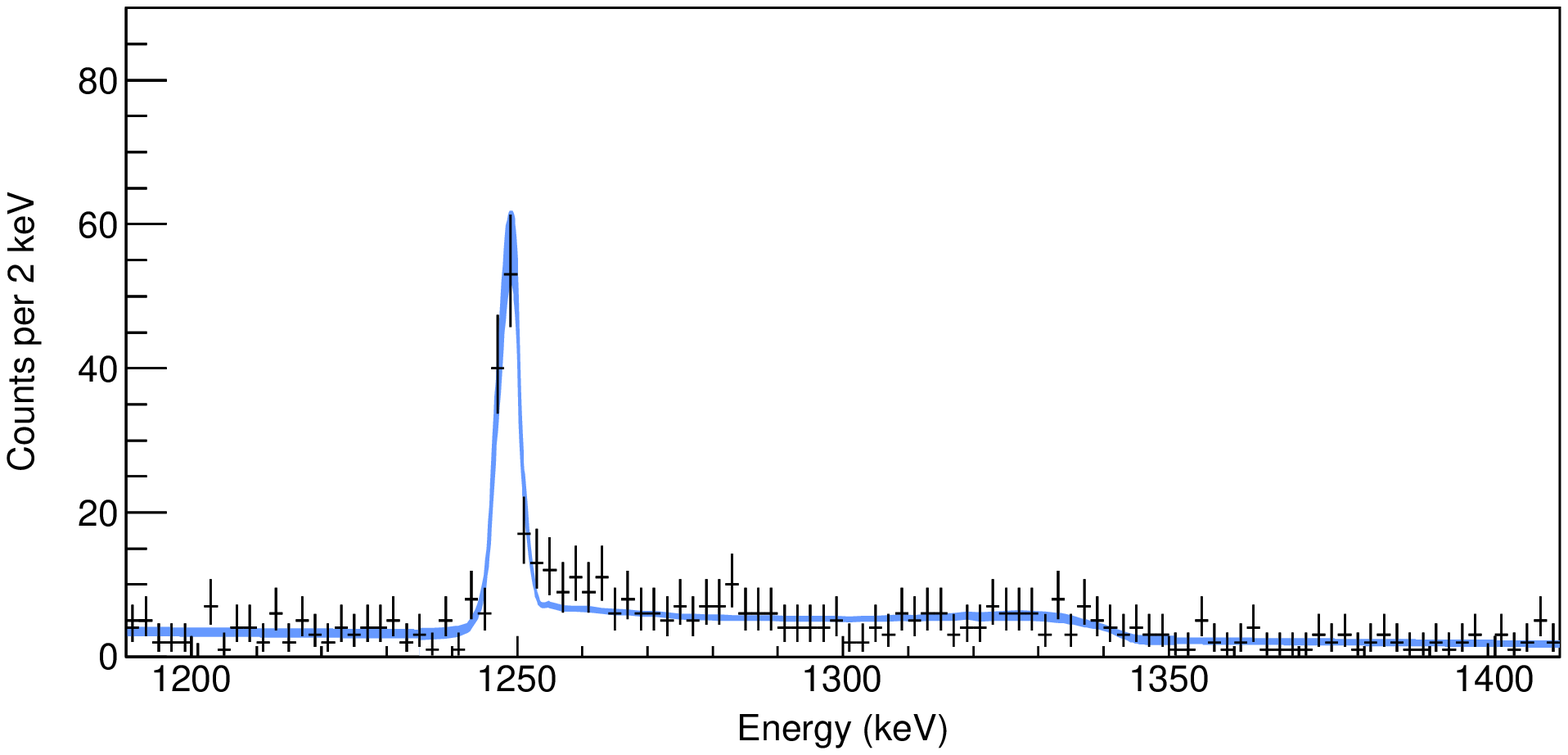}
\caption{\label{Gamma1248}\color{black}Lineshape analysis of the 49-MeV $\alpha$-gated $\gamma$-ray line from the 1248~keV $\rightarrow$ ground state transition in $^{31}$S. The measured lineshape is shown as points with statistical error bars in both panels. Prior lineshape (red, upper panel): hundreds of lineshapes generated by varying each parameter within its prior range in the DSAM simulation. Posterior lineshape (blue, lower panel): 1$\sigma$ confidence band constructed with the number of counts in each bin corresponding to the parameter posterior distributions.}
\end{center}
\end{figure}

Fitting the simulated $\gamma$-ray spectrum to the measured $\gamma$-ray spectrum in a given range yields the number of counts and the associated standard deviation ($\sigma$). We set a discovery threshold for statistical significance over the background-only hypotheses to be 5$\sigma$~\cite{Zyla_PTEP2020}. {\color{black}The observed $^{31}$S $\gamma$-ray peaks are shown in Fig.~\ref{Gamma1248} and Supplementary Figs.~2-7 with the prior and posterior lineshapes superimposed (See Sec.~\ref{Bayesian_Analyses}).} The gate on the energy deposited by the $\alpha$ particles is 2~MeV wide in all cases, corresponding to a 1-MeV window on the excitation energies. We used a fine binning of 2~keV in each lineshape analysis to mitigate the information loss associated with the bin size of the spectrum. The 1248-keV, 2234-keV, 3076-keV, 4971-keV, and 5156-keV $^{31}$S states all decay predominantly by a single $\gamma$-ray transition to the ground state~\cite{Bennett_PRC2018}, and their dominant $\gamma$ rays are clearly observed in the corresponding $\alpha$-gated $\gamma$-ray spectra. Other than these five peaks, a 2186-keV line from the decay of the 3435-keV state is also observed with a statistical significance greater than 5$\sigma$. There are two $\gamma$ rays which are emitted from the 3435-keV $^{31}$S state with branching ratios of 54.7(35)\% and 45.3(30)\% to the ground and first excited states, respectively~\cite{Bennett_PRC2018}. The observed significance of the 3435-keV line does not reach the 5$\sigma$ threshold, so we extract the lifetime based on the lineshape analysis of the 2186-keV line. We estimated the branching ratios to be 38(11)\% and 62(12)\% using the efficiency-corrected counts in the 3435- and 2186-keV $\gamma$-ray peaks with only statistical uncertainties included, consistent with the previous measurement~\cite{Bennett_PRC2018}. All six aforementioned $^{31}$S states were also observed to be populated in the $^{32}$S$(^{3}$He$,\alpha)^{31}$S reaction at the same center-of-mass energy as ours~\cite{Moss_NPA1970}.

$\gamma$-ray transitions from the $3/2^+$, 6390.2(7)-keV state at 2183, 3106, 3314, 4156, 5141, and 6390~keV were previously identified~\cite{Bennett_PRC2018}. Evidence for the 4156 and 5141-keV branches was observed in this data set with significances over 4$\sigma$ and 3$\sigma$, respectively. Consistent with past work~\cite{Bennett_PRC2018}, the 4156-keV branch is the strongest branch observed. {\color{black}A nearby $5/2^+$, 6392.5(2)-keV state was observed to dominantly populate the $3/2^+$, 1248-keV state with a 5143.1(2)-keV $\gamma$-ray branch~\cite{Doherty_PRL2012,Doherty_PRC2014,Kankainen_PLB2017}, which could be responsible for the higher statistics we observed for the 5141-keV branch. The absence of the $5/2^+$, 6392~keV $\rightarrow$ $5/2^+$, 2234~keV transition is also consistent with our shell model calculations.} Another nearby $11/2^+$, 6394.2(2)-keV state generates 1091.2(4)- and 3042.9(1)-keV branches~\cite{Doherty_PRL2012,Doherty_PRC2014,Jenkins_PRC2005,Jenkins_PRC2006,Testov_PRC2021}, but neither of these branches was observed in our spectra. A recent study using the $^{32}$S$(p,d)^{31}$S reaction channel also indicates that the $(3/2,5/2)^+$ states near 6390~keV are preferentially populated, but the $11/2^+$ state is not~\cite{Setoodehnia_PRC2020}. {\color{black}As the 4156-keV $\gamma$ ray is likely to be uniquely associated with the $3/2^+$, 6390-keV $\rightarrow$ $5/2^+$, 2234-keV transition, we attempt to extract the lifetime from its lineshape.}

\section{\label{Bayesian_Analyses}Bayesian Analyses}
\begin{figure}
\begin{center}
\includegraphics[width=8.6cm]{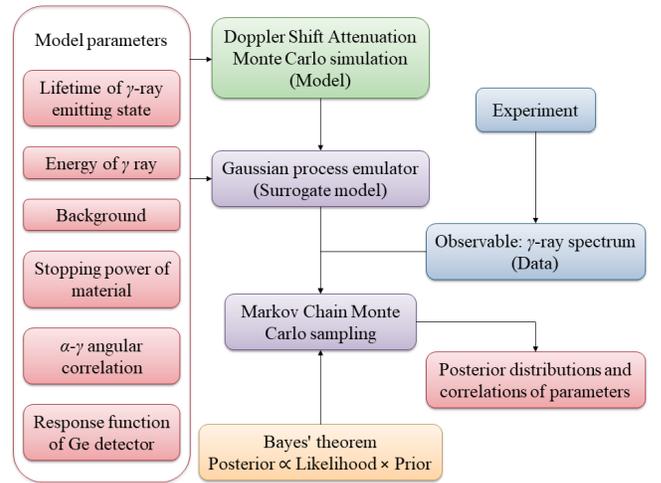}
\caption{\label{DSAM_workflow}Workflow of the \mbox{MCMC}-based Bayesian \mbox{DSAM} data analysis framework.}
\end{center}
\end{figure}

{\color{black}A graphical representation of our analysis procedure is shown in Fig.~\ref{DSAM_workflow}.}

In general, when using Bayes's theorem~\cite{Bayes_PT1763} to set up the problem of fitting a model to data, the procedure begins with a hypothesis, which is a set of model parameters to be estimated, $\boldsymbol{x}$, and a set of experimental data, $\boldsymbol{D}$, to be compared with model calculations. We then define a likelihood, $P(\boldsymbol{D}|\boldsymbol{x})$, the probability of the data $\boldsymbol{D}$ being observed given the parameters $\boldsymbol{x}$, which is determined by running the model with parameters $\boldsymbol{x}$ and fitting the model output to data. Likelihood quantifies how well the model reproduces the data. A prior probability distribution, $P(\boldsymbol{x})$, encapsulates our initial belief of the parameters. Next, we invoke Bayes's theorem, which states that our updated belief after observing the data, i.e., the posterior probability distribution of the model parameters given the data, $P(\boldsymbol{x}|\boldsymbol{D})$, is proportional to the product of the likelihood and the prior:

\begin{equation}
P(\boldsymbol{x}|\boldsymbol{D})= \frac{P(\boldsymbol{D}|\boldsymbol{x})P(\boldsymbol{x})}{P(\boldsymbol{D})}
\end{equation}

The denominator, $P(\boldsymbol{D})$, is the Bayesian evidence, which is the probability of observing the data without having compared to the model and, given that the data are known, serves as a normalization factor. Translating for the application here, our model is a $\mbox{DSAM}$ simulation convoluted with a linear fit function to describe the background, which introduces several parameters. The output of the model is a $\gamma$-ray spectrum, which is the observable to be compared with data. For the choice of priors $P(\boldsymbol{x})$, we specify ranges and distributions for each parameter. A uniform prior is chosen for the lifetime $\tau$. Negative lifetimes are unphysical, so we set the prior to be zero in negative regions:
\begin{equation}
P(\tau)=\left\{
\begin{array}{lr}
\mathrm{constant}, & \tau>0 \\
0, & \tau\leqslant0
\end{array}
\right.
\end{equation}

We use a Gaussian distribution for the $\gamma$-ray energy, $E_\gamma$, from the literature values and uncertainties~\cite{Bennett_PRC2018}. We construct a Gaussian distribution for the relative background level, $bkg$, based on the linear fit and its uncertainty in the background region around a $\gamma$-ray peak. The stopping power incorporated in \textsc{geant}4 is expected to be overall accurate to within 10\%~\cite{ICRU73_2005}. Accuracy is generally higher in the energy range above 10~MeV/nucleon, while the uncertainty increases at energies below 0.1~MeV/nucleon. We use a Gaussian distribution for the relative stopping power, $sp$, centered at the database values with a 1$\sigma$ uncertainty of 10\% for short-lived states and 20\% for long-lived states, respectively. The prior on the coefficient of the Legendre polynomial $P_2(\mathrm{cos}\theta$) of the $\alpha$-$\gamma$ angular-correlation function, $A_2$, is assumed to be uniform within $[-1,1]$, and $A_4$ is fixed to be 0. When a $\gamma$ ray is emitted from a long-lived state, the emission usually happens after the recoil has undergone a series of collisions with the target atoms. Hence, the $\gamma$-ray lineshape is quite insensitive to variations of the angular correlation function. We, therefore, omit the angular-correlation parameter for the two long-lived states. Our primary goal is to learn the unknown model parameter $\tau$ from observables. $\tau$ is the parameter of interest, and the other four, $E_\gamma$, $bkg$, $sp$, and $A_2$, are referred to as nuisance parameters, which we do not aim to constrain using this data set.

To perform a bin-by-bin analysis using Bayes's theorem~\cite{Rodgers_NIMA2021}, we take the conditional probability of acquiring a measured set of data given the parameters $\boldsymbol{x}$ to be the likelihood function $\mathcal{L}(\boldsymbol{x})$:

\begin{equation}
\mathcal{L}(\boldsymbol{x})\approx \mathrm{exp}\left[-\sum_{i=1}^{n}\frac{[y_i^{\mathrm{exp}}-y_i^{\mathrm{mod}}(\boldsymbol x)]^2}{2\sigma_i^2}\right],
\end{equation}

where $n$ is the number of bins, $y_i^{\mathrm{exp}}$ is the number of counts in the $i$th bin of the measured spectrum, and $y_i^{\mathrm{mod}}$ is the number of counts in the $i$th bin predicted by the model. $\sigma_i$ accounts for both experimental and theoretical uncertainties, including the emulator predictive uncertainty~\cite{Novak_PRC2014,Pratt_PRL2015}. We assume all the uncertainties are Gaussian.

\mbox{MCMC} algorithms generate a random walk through the parameter space where each step is accepted or rejected according to the product of the prior and the likelihood to reproduce the measured observables~\cite{Metropolis_JCP1953,Hastings_Bio1970}. Direct \mbox{MCMC} sampling requires millions of model evaluations. In our case, a single model evaluation requires thousands of individual event simulations and is computationally demanding, so direct \mbox{MCMC} sampling is intractable. As our model space is relatively low-dimensional, we choose a factorial design, in which hundreds of design points uniformly fill the parameter space like a grid. {\color{black}We run the full DSAM simulation at these design points, and the model outputs are transformed into a reduced number of uncorrelated variables using the principal component analysis~\cite{Tipping_NC1999}. A Gaussian Process (GP) emulator~\cite{Rasmussen_2006} is trained on the input-output behavior of the full model and acts as a fast surrogate to the full model during \mbox{MCMC} sampling. GP is computationally efficient and accurately accounts for the uncertainty associated with emulation, which is suited for Bayesian parameter estimation purposes.} The highest computational cost in the procedure is now associated with obtaining full-model data to train the GP emulator, which can usually be accomplished in a realistic amount of time.

The Modeling and Data Analysis Initiative (MADAI) collaboration~\cite{MADAI} developed a statistical framework that contains a GP emulator and a \mbox{MCMC} sampler. We have tailored the MADAI infrastructure to our needs. We explored a five-dimensional parameter space by discarding a 50,000-step burn-in phase for the chain to converge and then sampling for another one million \mbox{MCMC} steps. {\color{black}For all but the 4156-keV $\gamma$ ray, every individual MCMC chain was able to achieve adequate convergence to the posterior distribution with no more than a few thousand iterations.} We estimated the hyperparameters by numerically maximizing the likelihood but found that varying hyperparameters only weakly affects the actual emulator predictions.

\begin{figure}
\begin{center}
\includegraphics[width=8.6cm]{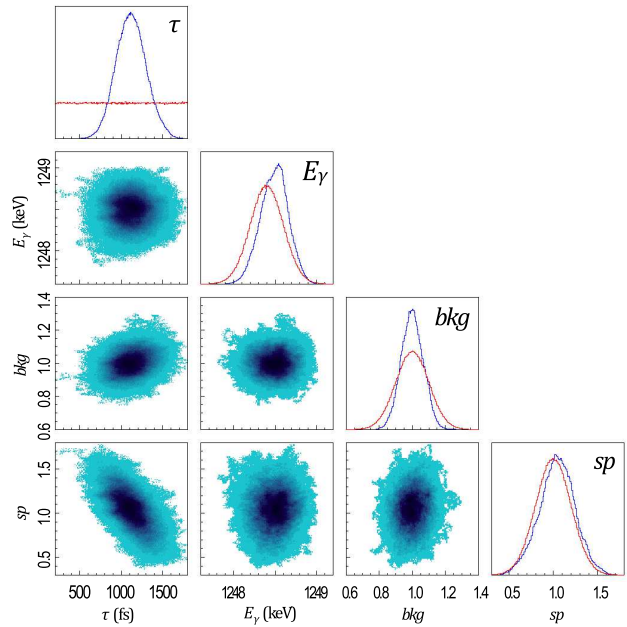}
\caption{\label{Posterior1248}Posterior distributions of the model parameters for the $^{31}$S $3/2^+$ state at 1248~keV. Diagonals: prior (red) and posterior (blue) distributions of each parameter. From top-left to bottom right: Lifetime $\tau$~(fs), $\gamma$-ray energy $E_\gamma$~(keV), relative background $bkg$, and relative stopping power $sp$. Off-diagonals: joint distributions showing correlations between pairs of parameters.}
\end{center}
\end{figure}

The diagonal panels in Fig.~\ref{Posterior1248} and Supplementary Figs.~8-13 show the marginal distributions for each parameter with all other parameters integrated out, and the off-diagonal panels show joint distributions between pairs of parameters. The prior distributions for the nuisance parameters, $E_\gamma$, $bkg$, and $sp$, are restrictive and strongly influence the posteriors. {\color{black}One important merit of Bayesian methods is that a posterior distribution offers more detailed information than a point estimate or an interval from frequentist methods so that propagation of uncertainty can work with richer information than that conveyed by a point estimate~\cite{Sharma_ARAA2017,JCGM_2008}. The 2D correlations between parameters allow us to easily capture features, patterns, or anomalies. Strong negative correlations are demonstrated between $\tau$ and $sp$ for the two long-lived states, which is physically expected. These two parameters are not correlated for short-lived states as the deexcitations occur before substantial slowing down of the recoils occurs. For the observable itself, the lineshapes based on the prior and posterior distributions of parameters are shown in Fig.~\ref{Gamma1248} and Supplementary Figs.~2-7. The fact that the posterior bands are narrow and closely resemble the measured lineshapes demonstrates the constraint provided by the experimental data.}


\section{Lifetimes}
For the two lowest-lying states, the central lifetime values and the 1$\sigma$ uncertainties are constructed by using the 16th, 50th, and 84th percentile values from the lifetime posterior distributions. For the four higher-lying states, the most probable lifetime values are close to zero, and therefore, the 90th percentile values for the lifetime posterior distributions are adopted as the 90\% confidence upper limits. For the first excited state at 1248~keV, we obtained a lifetime of $1120\pm180$~fs, which agrees with the literature values of $720\pm180$~fs~\cite{Engmann_NPA1971}, $1200^{+1500}_{-1100}$~fs~\cite{Doornenbal_NIMA2010}, $3200\pm7000$~fs~\cite{Doornenbal_NIMA2010}, and $964^{+312}_{-91}$~fs~\cite{Herlitzius_Thesis2013}. Tonev~\textit{et al}. recently reported a lifetime of $624\pm32$~fs~\cite{Tonev_PLB2021} for the 1248-keV state, lower than all the other results. They reported $543\pm49$~fs for the $7/2^-$ state at 4451~keV, which is also much lower than another measurement of $1030\pm210 $~fs~\cite{Pattabiraman_PRC2008}. For the second excited state at 2234~keV, our result $250\pm80$~fs agrees with the only literature value of $320\pm80$~fs~\cite{Engmann_NPA1971}. We obtained consistent lifetime values and limits for all $\gamma$-ray lines using the standard frequentist approach.

We performed theoretical calculations using the shell-model code \textsc{NuShellX}~\cite{Brown_NDS2014} in the $sd$-shell-model space involving the $\pi0d_{5/2}$, $\pi1s_{1/2}$, $\pi0d_{3/2}$, $\nu0d_{5/2}$, $\nu1s_{1/2}$, and $\nu0d_{3/2}$ valence orbits. Five universal $sd$-shell type A (USDA)~\cite{Brown_PRC2006}, type B (USDB)~\cite{Brown_PRC2006}, type C (USDC)~\cite{Magilligan_PRC2020}, type E (USDE), and type I (USDI)~\cite{Magilligan_PRC2020} Hamiltonians have been used in our calculations. Given that decay widths are very sensitive to energies, we have applied a correction to the theoretical $\gamma$-ray partial widths ($\Gamma_\gamma$) based on the experimental energies~\cite{Bennett_PRC2018}. Each theoretical $\Gamma_\gamma$ is obtained using the effective $M$1 and $E$2 transition operators~\cite{Richter_PRC2008} and then scaled for the $E^{2L+1}_\gamma$ energy dependence, where $L$ denotes the multipolarity of the radiation.

\begin{table*}
\small
\begin{center}
\caption{\label{AllLifetime}Lifetimes of the two lowest-lying $^{31}$S states and the 90\% confidence upper limits on the lifetimes of four higher-lying $^{31}$S states measured in the present work are listed in column 4. The spins and parities ($J^\pi$), excitation energies ($E_x$), and $\gamma$-ray energies ($E_\gamma$) of the dominant branch for each state are adopted from Ref.~\cite{Bennett_PRC2018}. The excitation energies and the lifetimes of $^{31}$P mirror states listed in the last two columns are adopted from Ref.~\cite{Ouellet_NDS2013}. All $E_x$ and $E_\gamma$ are rounded to the closest integer. A hyphen ($-$) is placed where the value is unavailable.}
\begin{tabular}{ccccccccccccccccc}
\hline
$J^\pi$ & $E_x(^{31}\mathrm{S})$~(keV) & $E_\gamma$~(keV) & $\tau_{\mathrm{exp}}$~(fs) & $\tau_{\mathrm{USDA}}$~(fs) & $\tau_{\mathrm{USDB}}$~(fs) & $\tau_{\mathrm{USDC}}$~(fs) & $\tau_{\mathrm{USDE}}$~(fs) & $\tau_{\mathrm{USDI}}$~(fs) & $E_x(^{31}\mathrm{P})$~(keV) & $\tau(^{31}\mathrm{P})$~(fs) \\
\hline
$3/2^+$ & 1248   & 1248 & $1120(180)$   & 1794 & 2633 & 2428 & 2735 & 2734 & 1266 & 754(26) \\
$5/2^+$ & 2234   & 2234 & $250(80)$  & 285 & 311 & 306 & 325 & 317 & 2234 & 388(26) \\
$1/2^+$ & 3076   & 3076 & $<$11   & 16 & 14 & 13 & 15 & 12 & 3134 & 10.4(9) \\
$3/2^+$ & 3435   & 2186 & $<$16   & 19 & 16 & 15 & 15 & 14 & 3506 & 12.7(19) \\
$3/2^-$  & 4971   & 4970 & $<$7    & $-$ & $-$ & $-$ & $-$ & $-$ & 5015 & 11.0(7) \\
$1/2^+$ & 5156   & 5156 & $<$15  & 3.0 & 3.2 & 3.3 & 2.7 & 4.2 & 5257 & $<$15 \\
\hline
\end{tabular}
\end{center}
\end{table*}

All the measured and calculated lifetimes of $^{31}$S states are summarized in Table~\ref{AllLifetime}. The negative-parity 4971-keV state is not matched with any theoretical state as cross-shell excitations were not taken into account in our shell-model predictions. The calculated $\gamma$ decay of the 1248-keV state is dominated by an $M1$ transition. Correcting for the USDB-calculated partial lifetime for the $E2$ transition of 6.9~ps, the experimental partial lifetime for the $M1$ transition is 1.33(26)~ps. This gives an experimental transition probability of $B(M1)_{\mathrm{exp}}=0.022(4)~\mu_N^2$. With $B(M1)=[M(M1)]^2/(2J_i+1)$, where $M(M1)$ is the transition matrix element and $J_i$ is the spin of the $\gamma$-emitting state, we have $|M(M1)|_{\mathrm{exp}}=0.30(3)~\mu_N$ to be compared with, for example, $|M(M1)|_{\mathrm{USDA}}=0.22~\mu_N$ and $|M(M1)|_{\mathrm{USDB}}=0.17~\mu_N$. The comparison between theory and experiment for other $M(M1)$ values in the $sd$ shell is shown in Fig.~4 of Ref.~\cite{Richter_PRC2008}. It is observed that theory and experiment differ by about $\pm0.3~\mu_N$ independent of the size of $M(M1)$. The present results are consistent with this observation. The measured lifetimes of all other states are in good agreement with our shell-model calculations.
The lifetimes of most states in the mirror nucleus $^{31}$P have been well measured~\cite{Ouellet_NDS2013} and are listed in Table~\ref{AllLifetime} for comparison. The lifetimes for all the mirror states are consistent with isospin being a good symmetry in the $^{31}$P-$^{31}$S system.

Dedicated shell model calculations have been performed to reproduce the strong isospin mixing between the 6390-keV state and the nearby isobaric analog state~\cite{Bennett_PRL2016,Budner_PRL2022,Brown_PRC2014}. We obtained $\tau=1.3$~fs using a shifted USDC Hamiltonian. Limited mainly by the low statistics collected on the 6390 $\rightarrow$ 2234~keV transition, we are not able to set a finite constraint on the lifetime of the 6390-keV $^{31}$S state. The posterior clearly favors a short lifetime as it exceeds the prior below 20~fs (Supplementary Fig.~13). An upper limit of the lifetime $\tau<20$~fs is equivalent to a lower limit on the decay width of $\Gamma>33$~meV. Combining with the finite proton branching ratio value~\cite{Budner_PRL2022} yields a resonance strength of $\omega\gamma>5.5~\mu$eV, consistent with the previous $\omega\gamma=80(48)~\mu$eV based on the measured proton branching ratio and a theoretical lifetime~\cite{Budner_PRL2022}.

\begin{figure}
\begin{center}
\includegraphics[width=8.6cm]{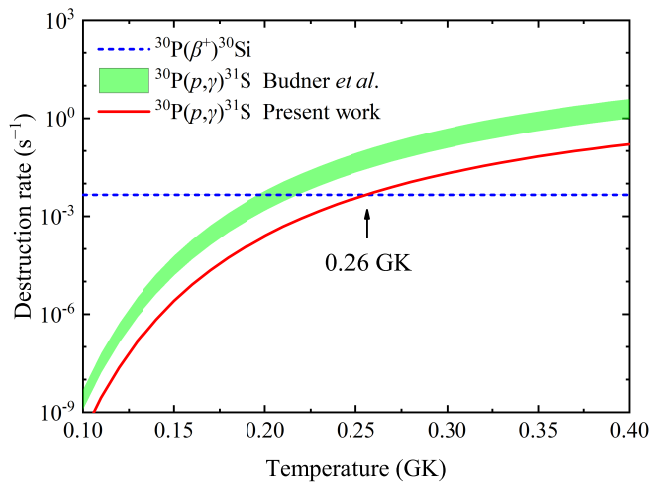}
\caption{\label{Decayconstant}$^{30}$P destruction rates for the $^{30}$P$(p,\gamma)^{31}$S reaction (solid red line) and $^{30}$P$(\beta^+)^{30}$Si decay (dashed blue line) as a function of temperature. Only the resonant-capture contribution from the $3/2^+$ resonance is taken into account, representing the lower limit of the $^{30}$P$(p,\gamma)^{31}$S reaction. The rate derived from Ref.~\cite{Budner_PRL2022} (green band) is shown for comparison.}
\end{center}
\end{figure}

{\color{black}Here we provide a qualitative picture of the astrophysical impact. The $^{30}$P$(p,\gamma)^{31}$S reaction and the $^{30}$P$(\beta^+)^{30}$Si decay are the two main destruction mechanisms for $^{30}$P in ONe novae~\cite{Jose_2016}. Assuming a stellar density of $\rho=300~\mathrm{g/cm}^3$ and a hydrogen mass fraction of $X_\mathrm{H}=0.3$~\cite{Iliadis_2015}, we derive the destruction rates for both processes from the literature $^{30}$P half-life of $T_{1/2}=2.498(5)$~min~\cite{Basunia_NDS2010} and the newly-determined lower limit on the strength of the $3/2^+$ resonance, $\omega\gamma=5.5~\mu$eV. Figure~\ref{Decayconstant} shows equal destruction rates of the two processes at 0.26~GK, implying that the proton capture becomes more likely than the competing $\beta^+$ decay beyond a temperature within the peak nova temperatures of $T_\mathrm{peak}=0.1$$-$0.4~GK. The location of the crossing point would affect interesting nova observables, such as the $^{30}$Si/$^{28}$Si isotopic abundance ratios useful for the identification of pre-solar nova grains~\cite{Jose_APJ2004}, the O/S, S/Al, O/P, and P/Al abundance ratios that are good candidates for nova thermometers~\cite{Downen_APJ2013}, and the Si/H abundance ratio as a useful nuclear mixing meter in ONe novae~\cite{Kelly_APJ2013}.}

\section{Conclusion \& Outlook}
{\color{black}To summarize, we performed $\mbox{DSAM}$ lifetime measurements of  $^{31}$S states using the DSL facility. We applied the MCMC-based Bayesian method to rigorously constrain model parameters and quantify uncertainties, demonstrating the usefulness of Bayesian parameter estimation for $\mbox{DSAM}$ lineshape analyses. As more powerful Bayesian tools are continuously being developed~\cite{Phillips_JPG2021,Bedaque_EPJA2021,Boehnlein_RMP2022}, we expect to see that the framework established in this work has broad applicability to more lineshape analyses.}

Our newly-determined lifetime upper limits for the four high-lying states contribute to the understanding of the nuclear structure of $^{31}$S. The observation of $\gamma$ rays from the 6390-keV state is very promising for future measurements with higher statistics. This work represents a major step toward an entirely experimentally-determined thermonuclear rate of the $^{30}$P$(p,\gamma)^{31}$S reaction. Advancing this work will be the DSL2 facility, consisting of a segmented Si detector telescope with higher solid angle coverage and reduced $\gamma$-ray attenuation. The granularity of the new telescope provides the position resolution necessary to maintain the angular/kinematic resolution that enables gating on excitation energies. The lifetime sensitivity will benefit greatly from the large solid angle and position resolution of the new telescope.

\section{Acknowledgements}
We gratefully acknowledge the TRIUMF staff for technical assistance and for providing the $^{32}$S beam. We would like to thank Pablo Giuliani, Caleb Marshall, Scott Pratt, Cole Pruitt, Chun Yuen Tsang, and Lei Yang for helpful discussions on Bayesian analyses. This work was supported by the U.S. National Science Foundation under Grants Nos. PHY-1102511, PHY-1565546, and PHY-2110365, and the U.S. Department of Energy, Office of Science, under Award No. DE-SC0016052. The authors acknowledge the generous support of the Natural Sciences and Engineering Research Council of Canada. TRIUMF receives federal funding via a contribution agreement through the National Research Council of Canada.


\begin{thebibliography}{400}

\bibitem{Jose_2016}
J. Jos\'{e}, \textit{Stellar Explosions: Hydrodynamics and Nucleosynthesis} (Boca Raton, FL: CRC Press, 2016).

\bibitem{Iliadis_2015}
C. Iliadis, \textit{Nuclear Physics of Stars} (Wiley-VCH, Verlag, Wenheim, Germany, 2015).

\bibitem{Jose_NPA2006}
J. Jos\'{e}, M. Hernanz, and C. Iliadis, \href{https://doi.org/10.1016/j.nuclphysa.2005.02.121}{Nucl. Phys. A \textbf{777}, 550 (2006).}

\bibitem{Jose_APJ2004}
J. Jos\'{e}, M. Hernanz, S. Amari, K. Lodders, and E. Zinner, \href{https://doi.org/10.1086/422569}{Astrophys. J. \textbf{612}, 414 (2004).}

\bibitem{Downen_APJ2013}
Lori N. Downen, Christian Iliadis, Jordi Jos\'{e}, and Sumner Starrfield, \href{https://doi.org/10.1088/0004-637x/762/2/105}{Astrophys. J. \textbf{762}, 105 (2013).}

\bibitem{Kelly_APJ2013}
Keegan J. Kelly, Christian Iliadis, Lori Downen, Jordi Jos\'{e}, and Art Champagne, \href{https://doi.org/10.1088/0004-637x/777/2/130}{Astrophys. J. \textbf{777}, 130 (2013).}

\bibitem{Wrede_AIPA2014}
C. Wrede, \href{https://doi.org/10.1063/1.4864193}{AIP Advances \textbf{4}, 041004 (2014).}

\bibitem{Bennett_PRL2016}
M. B. Bennett, C. Wrede, B. A. Brown, S. N. Liddick, D. P\'{e}rez-Loureiro, D. W. Bardayan, A. A. Chen, K. A. Chipps, C. Fry, B. E. Glassman, C. Langer, N. R. Larson, E. I. McNeice, Z. Meisel, W. Ong, P. D. O'Malley, S. D. Pain, C. J. Prokop, H. Schatz, S. B. Schwartz, S. Suchyta, P. Thompson, M. Walters, and X. Xu, \href{https://doi.org/10.1103/PhysRevLett.116.102502}{Phys. Rev. Lett. \textbf{116}, 102502 (2016).}

\bibitem{Bennett_PRC2018}
M. B. Bennett, C. Wrede, S. N. Liddick, D. P\'{e}rez-Loureiro, D. W. Bardayan, B. A. Brown, A. A. Chen, K. A. Chipps, C. Fry, B. E. Glassman, C. Langer, N. R. Larson, E. I. McNeice, Z. Meisel, W. Ong, P. D. O'Malley, S. D. Pain, C. J. Prokop, H. Schatz, S. B. Schwartz, S. Suchyta, P. Thompson, M. Walters, and X. Xu, \href{https://doi.org/10.1103/PhysRevC.97.065803}{Phys. Rev. C \textbf{97}, 065803 (2018).}

\bibitem{Budner_PRL2022}
T. Budner, M. Friedman, C. Wrede, B. A. Brown, J. Jos\'{e}, D. P\'{e}rez-Loureiro, L. J. Sun, J. Surbrook, Y. Ayyad, D. Bardayan, K. Chae, A. Chen, K. Chipps, M. Cortesi, B. Glassman, M. R. Hall, M. Janasik, J. Liang, P. O'Malley, E. Pollacco, A. Psaltis, J. Stomps, and T. Wheeler, \href{https://doi.org/10.1103/PhysRevLett.128.182701}{Phys. Rev. Lett. \textbf{128}, 182701 (2022).}

\bibitem{Engmann_NPA1971}
R. Engmann, E. Ehrmann, F. Brandolini, and C. Signorini, \href{https://doi.org/10.1016/0375-9474(71)90986-9}{Nucl. Phys. A \textbf{162}, 295 (1971).}

\bibitem{Doornenbal_NIMA2010}
P. Doornenbal, P. Reiter, H. Grawe, T. Saito, A. Al-Khatib, A. Banu, T. Beck, F. Becker, P. Bednarczyk, G. Benzoni, A. Bracco, A. B\"{u}rger, L. Caceres, F. Camera, S. Chmel, F.C.L. Crespi, H. Geissel, J. Gerl, M. G\'{o}rska, J. Gre,bosz, H. H\"{u}bel, M. Kavatsyuk, O. Kavatsyuk, M. Kmiecik, I. Kojouharov, N. Kurz, R. Lozeva, A. Maj, S. Mandal, W. Meczynski, B. Million, Zs. Podoly\'{a}k, A. Richard, N. Saito, H. Schaffner, M. Seidlitz, T. Striepling, J. Walker, N. Warr, H. Weick, O. Wieland, M. Winkler, and H.J. Wollersheim, \href{https://doi.org/10.1016/j.nima.2009.11.017}{Nucl. Instrum. Methods Phys. Res. A \textbf{613}, 218 (2010).}

\bibitem{Herlitzius_Thesis2013}
Clemens Herlitzius, \href{http://mediatum.ub.tum.de/?id=1145824}{Ph.D. Thesis}, Technische Universit\"{a}t M\"{u}nchen, Germany, 2013.

\bibitem{Tonev_PLB2021}
D. Tonev, G. de Angelis, I. Deloncle, N. Goutev, G. De Gregorio, P. Pavlov, I.L. Pantaleev, S. Iliev, M.S. Yavahchova, P.G. Bizzeti, A. Demerdjiev, D.T. Dimitrov, E. Farnea, A. Gadea, E. Geleva, C.Y. He, H. Laftchiev, S.M. Lenzi, S. Lunardi, N. Marginean, R. Menegazzo, D.R. Napoli, F. Nowacki, R. Orlandi, H. Penttil\"{a}, F. Recchia, E. Sahin, R.P. Singh, M. Stoyanova, C.A. Ur, H.-F. Wirth, \href{https://doi.org/10.1016/j.physletb.2021.136603}{Phys. Lett. B \textbf{821}, 136603 (2021).}

\bibitem{Pattabiraman_PRC2008}
N. S. Pattabiraman, D. G. Jenkins, M. A. Bentley, R. Wadsworth, C. J. Lister, M. P. Carpenter, R. V. F. Janssens, T. L. Khoo, T. Lauritsen, D. Seweryniak, S. Zhu, G. Lotay, P. J. Woods, Krishichayan, and P. Van Isacker, \href{https://doi.org/10.1103/PhysRevC.78.024301}{Phys. Rev. C \textbf{78}, 024301 (2008).}

\bibitem{Elliott_PR1948}
L. G. Elliott and R. E. Bell, \href{https://doi.org/10.1103/PhysRev.74.1869}{Phys. Rev. \textbf{74}, 1869 (1948).}

\bibitem{Devons_PPSA1955}
S. Devons, G. Manning and D. St. P. Bunbury, \href{https://doi.org/10.1088/0370-1298/68/1/304}{Proc. Phys. Soc. A \textbf{68}, 18 (1955).}

\bibitem{Devons_PPSA1956}
S. Devons, G. Manning, and J. H. Towle, \href{https://doi.org/10.1088/0370-1298/69/2/311}{Proc. Phys. Soc. A \textbf{69}, 173 (1956).}

\bibitem{King_PRL2019}
G. B. King, A. E. Lovell, L. Neufcourt, and F. M. Nunes, \href{https://doi.org/10.1103/PhysRevLett.122.232502}{Phys. Rev. Lett. \textbf{122}, 232502 (2019).}

\bibitem{Catacora-Rios_PRC2021}
M. Catacora-Rios, G. B. King, A. E. Lovell, and F. M. Nunes, \href{https://doi.org/10.1103/PhysRevC.104.064611}{Phys. Rev. C \textbf{104}, 064611 (2021).}

\bibitem{Lovell_JPG2021}
A. E. Lovell, F. M. Nunes, M. Catacora-Rios, and G. B. King, \href{https://doi.org/10.1088/1361-6471/abba72}{J. Phys. G: Nucl. Part. Phys. \textbf{48}, 014001 (2021).}

\bibitem{Colin_2013}
Fox Colin, Haario Heikki, Christen J Andr\'{e}s, \textit{Inverse problems} in Paul Damien and others (eds), \href{https://doi.org/10.1093/acprof:oso/9780199695607.001.0001}{\textit{Bayesian Theory and Applications} (Oxford University Press, 2013).}

\bibitem{Bayes_PT1763}
Thomas Bayes and Richard Price, \href{https://doi.org/10.1098/rstl.1763.0053}{Philos. Trans. \textbf{53}, 370 (1763).}

\bibitem{Jeffreys_1939}
H. Jeffreys, \textit{The Theory of Probability} (Oxford, UK: Oxford Univ. Press, 1939).

\bibitem{Metropolis_JCP1953}
N. Metropolis, A. W. Rosenbluth, M. N. Rosenbluth, A. H. Teller, and E. Teller, \href{https://doi.org/10.1063/1.1699114}{J. Chem. Phys. \textbf{21}, 1087 (1953).}

\bibitem{Hastings_Bio1970}
W. K. Hastings, \href{https://doi.org/10.1093/biomet/57.1.97}{Biometrika \textbf{57}, 97 (1970).}

\bibitem{Sharma_ARAA2017}
Sanjib Sharma, \href{https://doi.org/10.1146/annurev-astro-082214-122339}{Annu. Rev. Astron. Astrophys. \textbf{55}, 213 (2017).}

\bibitem{Schoot_NRMP2021}
Rens van de Schoot, Sarah Depaoli, Ruth King, Bianca Kramer, Kaspar M\"{a}rtens, Mahlet G. Tadesse, Marina Vannucci, Andrew Gelman, Duco Veen, Joukje Willemsen, and Christopher Yau, \href{https://doi.org/10.1038/s43586-020-00001-2}{Nat. Rev. Methods Primers \textbf{1}, 1 (2021).}



\bibitem{Novak_PRC2014}
J. Novak, K. Novak, S. Pratt, J. Vredevoogd, C. E. Coleman-Smith, and R. L. Wolpert, \href{https://doi.org/10.1103/PhysRevC.89.034917}{Phys. Rev. C \textbf{89}, 034917 (2014).}

\bibitem{Pratt_PRL2015}
Scott Pratt, Evan Sangaline, Paul Sorensen, and Hui Wang, \href{https://doi.org/10.1103/PhysRevLett.114.202301}{Phys. Rev. Lett. \textbf{114}, 202301 (2015).}


\bibitem{Bernhard_PRC2016}
Jonah E. Bernhard, J. Scott Moreland, Steffen A. Bass, Jia Liu, and Ulrich Heinz, \href{https://doi.org/10.1103/PhysRevC.94.024907}{Phys. Rev. C \textbf{94}, 024907 (2016).}


\bibitem{Bernhard_NP2019}
Jonah E. Bernhard, J. Scott Moreland, and Steffen A. Bass, \href{https://doi.org/10.1038/s41567-019-0611-8}{Nat. Phys. \textbf{15}, 1113 (2019).}

\bibitem{Morfouace_PLB2019}
P. Morfouace, C.Y. Tsang, Y. Zhang, W.G. Lynch, M.B. Tsang, D.D.S. Coupland, M. Youngs, Z. Chajecki, M.A. Famiano, T.K. Ghosh, G. Jhang, Jenny Lee, H. Liu, A. Sanetullaev, R. Showalter, and J. Winkelbauer, \href{https://doi.org/10.1016/j.physletb.2019.135045}{Phys. Lett. B \textbf{799}, 135045 (2019).}

\bibitem{Moreland_PRC2020}
J. Scott Moreland, Jonah E. Bernhard, and Steffen A. Bass, \href{https://doi.org/10.1103/PhysRevC.101.024911}{Phys. Rev. C \textbf{101}, 024911 (2020).}

\bibitem{Xie_JPG2021}
Wen-Jie Xie and Bao-An Li, \href{https://doi.org/10.1088/1361-6471/abd25a}{J. Phys. G: Nucl. Part. Phys. \textbf{48}, 025110 (2021).}

\bibitem{Lovell_PRC2018}
A. E. Lovell and F. M. Nunes, \href{https://doi.org/10.1103/PhysRevC.97.064612}{Phys. Rev. C \textbf{97}, 064612 (2018).}

\bibitem{Yang_PLB2020}
L. Yang, C.J. Lin, Y.X. Zhang, P.W. Wen, H.M. Jia, D.X. Wang, N.R. Ma, F. Yang, F.P. Zhong, S.H. Zhong, T.P. Luo, \href{https://doi.org/10.1016/j.physletb.2020.135540}{Phys. Lett. B \textbf{807}, 135540 (2020).}

\bibitem{Marshall_PRC2020}
C. Marshall, P. Morfouace, N. de S\'{e}r\'{e}ville, and R. Longland, \href{https://doi.org/10.1103/PhysRevC.102.024609}{Phys. Rev. C \textbf{102}, 024609 (2020).}

\bibitem{Pruitt_PRC2020}
C. D. Pruitt, R. J. Charity, L. G. Sobotka, J. M. Elson, D. E. M. Hoff, K. W. Brown, M. C. Atkinson, W. H. Dickhoff, H. Y. Lee, M. Devlin, N. Fotiades, and S. Mosby, \href{https://doi.org/10.1103/PhysRevC.102.034601}{Phys. Rev. C \textbf{102}, 034601 (2020).}

\bibitem{Iliadis_APJ2016}
C. Iliadis, K. S. Anderson, A. Coc, F. X. Timmes, and S. Starrfield, \href{https://doi.org/10.3847/0004-637X/831/1/107}{Astrophys. J. \textbf{831}, 107 (2016).}

\bibitem{Zhang_Nature2022}
Liyong Zhang, Jianjun He, Richard J. deBoer, Michael Wiescher, Alexander Heger, Daid Kahl, Jun Su, Daniel Odell, Yinji Chen, Xinyue Li, Jianguo Wang, Long Zhang, Fuqiang Cao, Hao Zhang, Zhicheng Zhang, Xinzhi Jiang, Luohuan Wang, Ziming Li, Luyang Song, Hongwei Zhao, Liangting Sun, Qi Wu, Jiaqing Li, Baoqun Cui, Lihua Chen, Ruigang Ma, Ertao Li, Gang Lian, Yaode Sheng, Zhihong Li, Bing Guo, Xiaohong Zhou, Yuhu Zhang, Hushan Xu, Jianping Cheng, and Weiping Liu, \href{https://doi.org/10.1038/s41586-022-05230-x}{Nature \textbf{610}, 656 (2022).}

\bibitem{Phillips_JPG2021}
D. R. Phillips, R. J. Furnstahl, U. Heinz, T. Maiti, W. Nazarewicz, F. M. Nunes, M. Plumlee, M. T. Pratola, S. Pratt, F. G. Viens, and S. M. Wild, \href{https://doi.org/10.1088/1361-6471/abf1df}{J. Phys. G: Nucl. Part. Phys. \textbf{48}, 072001 (2021).}

\bibitem{Bedaque_EPJA2021}
Paulo Bedaque, Amber Boehnlein, Mario Cromaz, Markus Diefenthaler, Latifa Elouadrhiri, Tanja Horn, Michelle Kuchera, David Lawrence, Dean Lee, Steven Lidia, Robert McKeown, Wally Melnitchouk, Witold Nazarewicz, Kostas Orginos, Yves Roblin, Michael Scott Smith, Malachi Schram, Xin-Nian Wang, \href{https://doi.org/10.1140/epja/s10050-020-00290-x}{Eur. Phys. J. A \textbf{57}, 100 (2021).}

\bibitem{Boehnlein_RMP2022}
Amber Boehnlein, Markus Diefenthaler, Nobuo Sato, Malachi Schram, Veronique Ziegler, Cristiano Fanelli, Morten Hjorth-Jensen, Tanja Horn, Michelle P. Kuchera, Dean Lee, Witold Nazarewicz, Peter Ostroumov, Kostas Orginos, Alan Poon, Xin-Nian Wang, Alexander Scheinker, Michael S. Smith, and Long-Gang Pang, \href{https://doi.org/10.1140/epja/s10050-020-00290-x}{Rev. Mod. Phys. \textbf{94}, 031003 (2022).}

\bibitem{Galinski_PRC2014}
N. Galinski, S. K. L. Sjue, G. C. Ball, D. S. Cross, B. Davids, H. Al Falou, A. B. Garnsworthy, G. Hackman, U. Hager, D. A. Howell, M. Jones, R. Kanungo, R. Kshetri, K. G. Leach, J. R. Leslie, M. Moukaddam, J. N. Orce, E. T. Rand, C. Ruiz, G. Ruprecht, M. A. Schumaker, C. E. Svensson, S. Triambak, and C. D. Unsworth, \href{https://doi.org/10.1103/PhysRevC.90.035803}{Phys. Rev. C \textbf{90}, 035803 (2014).}

\bibitem{Davids_HI2014}
B. Davids, \href{https://doi.org/10.1007/s10751-013-0900-z}{Hyperfine Interact. \textbf{225}, 215 (2014).}

\bibitem{Kanungo_PRC2006}
R. Kanungo, T. K. Alexander, A. N. Andreyev, G. C. Ball, R. S. Chakrawarthy, M. Chicoine, R. Churchman, B. Davids, J. S. Forster, S. Gujrathi, G. Hackman, D. Howell, J. R. Leslie, A. C. Morton, S. Mythili, C. J. Pearson, J. J. Ressler, C. Ruiz, H. Savajols, M. A. Schumaker, I. Tanihata, P. Walden, and S. Yen, \href{https://doi.org/10.1103/PhysRevC.74.045803}{Phys. Rev. C \textbf{74}, 045803 (2006).}

\bibitem{Mythili_PRC2008}
S. Mythili, B. Davids, T. K. Alexander, G. C. Ball, M. Chicoine, R. S. Chakrawarthy, R. Churchman, J. S. Forster, S. Gujrathi, G. Hackman, D. Howell, R. Kanungo, J. R. Leslie, E. Padilla, C. J. Pearson, C. Ruiz, G. Ruprecht, M. A. Schumaker, I. Tanihata, C. Vockenhuber, P. Walden, and S. Yen, \href{https://doi.org/10.1103/PhysRevC.77.035803}{Phys. Rev. C \textbf{77}, 035803 (2008).}

\bibitem{Kirsebom_PRC2016}
O. S. Kirsebom, P. Bender, A. Cheeseman, G. Christian, R. Churchman, D. S. Cross, B. Davids, L. J. Evitts, J. Fallis, N. Galinski, A. B. Garnsworthy, G. Hackman, J. Lighthall, S. Ketelhut, P. Machule, D. Miller, S. T. Nielsen, C. R. Nobs, C. J. Pearson, M. M. Rajabali, A. J. Radich, A. Rojas, C. Ruiz, A. Sanetullaev, C. D. Unsworth, and C. Wrede, \href{https://doi.org/10.1103/PhysRevC.93.025802}{Phys. Rev. C \textbf{93}, 025802 (2016).}

\bibitem{ORTEC_B}
\href{https://www.ortec-online.com/products/radiation-detectors/silicon-charged-particle-radiation-detectors/si-charged-particle-radiation-detectors-for-research-applications/b-series}{ORTEC B Series Totally Depleted Silicon Surface Barrier Radiation Detector}.

\bibitem{Rizwan_NIMA2016}
U. Rizwan, A. B. Garnsworthy, C. Andreoiu, G. C. Ball, A. Chester, T. Domingo, R. Dunlop, G. Hackman, E. T. Rand, J. K. Smith, K. Starosta, C. E. Svensson, P. Voss, and J. Williams, \href{https://doi.org/10.1016/j.nima.2016.03.016}{Nucl. Instrum. Methods Phys. Res. A \textbf{820}, 126 (2016).}

\bibitem{Garnsworthy_NIMA2019}
A.B. Garnsworthy, C.E. Svensson, M. Bowry, R. Dunlop, A.D. MacLean, B. Olaizola, J.K. Smith, F.A. Ali, C. Andreoiu, J.E. Ash, W.H. Ashfield, G.C. Ball, T. Ballast, C. Bartlett, Z. Beadle, P.C. Bender, N. Bernier, S.S. Bhattacharjee, H. Bidaman, V. Bildstein, D. Bishop, P. Boubel, R. Braid, D. Brennan, T. Bruhn, C. Burbadge, A. Cheeseman, A. Chester, R. Churchman, S. Ciccone, R. Caballero-Folch, D.S. Cross, S. Cruz, B. Davids, A. Diaz Varela, I. Dillmann, M.R. Dunlop, L.J. Evitts, F.H. Garcia, P.E. Garrett, S. Georges, S. Gillespie, R. Gudapati, G. Hackman, B. Hadinia, S. Hallam, J. Henderson, S.V. Ilyushkin, B. Jigmeddorj, A.I. Kilic, D. Kisliuk, R. Kokke, K. Kuhn, R. Kr\"{u}cken, M. Kuwabara, A.T. Laffoley, R. Lafleur, K.G. Leach, J.R. Leslie, Y. Linn, C. Lim, E. MacConnachie, A.R. Mathews, E. McGee, J. Measures, D. Miller, W.J. Mills, W. Moore, D. Morris, L.N. Morrison, M. Moukaddam, C.R. Natzke, K. Ortner, E. Padilla-Rodal, O. Paetkau, J. Park, H.P. Patel, C.J. Pearson, E. Peters, E.E. Peters, J.L. Pore, A.J. Radich, M.M. Rajabali, E.T. Rand, K. Raymond, U. Rizwan, P. Ruotsalainen, Y. Saito, F. Sarazin, B. Shaw, J. Smallcombe, D. Southall, K. Starosta, M. Ticu, E. Timakova, J. Turko, R. Umashankar, C. Unsworth, Z.M. Wang, K. Whitmore, S. Wong, S.W. Yates, E.F. Zganjar, and T. Zidar, \href{https://doi.org/10.1016/j.nima.2018.11.115}{Nucl. Instrum. Methods Phys. Res. A \textbf{918}, 9 (2019).}

\bibitem{Ziegler_NIMB2010}
J. F. Ziegler, M. D.Ziegler, and J. P. Biersack, \href{https://doi.org/10.1016/j.nimb.2010.02.091}{Nucl. Instrum. Methods Phys. Res. B \textbf{268}, 1818 (2010).}

\bibitem{Huang_NDS2005}
Xiaolong Huang and Chunmei Zhou, \href{https://doi.org/10.1016/j.nds.2005.01.001}{Nucl. Data Sheets \textbf{104}, 283 (2005).}

\bibitem{Chen_NDS2018}
J. Chen, \href{https://doi.org/10.1016/j.nds.2018.03.001}{Nucl. Data Sheets \textbf{149}, 1 (2018).}

\bibitem{Tilley_NPA1993}
D.R. Tilley, H.R. Weller, and C.M. Cheves, \href{https://doi.org/10.1016/0375-9474(93)90073-7}{Nucl. Phys. A \textbf{564}, 1 (1993).}

\bibitem{Agostinelli_NIMA2003}
S. Agostinelli, J. Allison, K. Amako, J. Apostolakis, H. Araujo, P. Arce, M. Asai, D. Axen, S. Banerjee, G. Barrand, F. Behner, L. Bellagamba, J. Boudreau, L. Broglia, A. Brunengo, H. Burkhardt, S. Chauvie, J. Chuma, R. Chytracek, G. Cooperman, G. Cosmo, P. Degtyarenko, A. Dell'Acqua, G. Depaola, D. Dietrich, R. Enami, A. Feliciello, C. Ferguson, H. Fesefeldt, G. Folger, F. Foppiano, A. Forti, S. Garelli, S. Giani, R. Giannitrapani, D. Gibin, J. J. G\'{o}mez Cadenas, I. Gonz\'{a}lez, G. Gracia Abril, G. Greeniaus, W. Greiner, V. Grichine, A. Grossheim, S. Guatelli, P. Gumplinger, R. Hamatsu, K. Hashimoto, H. Hasui, A. Heikkinen, A. Howard, V. Ivanchenko, A. Johnson, F. W. Jones, J. Kallenbach, N. Kanaya, M. Kawabata, Y. Kawabata, M. Kawaguti, S. Kelner, P. Kent, A. Kimura, T. Kodama, R. Kokoulin, M. Kossov, H. Kurashige, E. Lamanna, T. Lamp\'{e}n, V. Lara, V. Lefebure, F. Lei, M. Liendl, W. Lockman, F. Longo, S. Magni, M. Maire, E. Medernach, K. Minamimoto, P. Mora de Freitas, Y. Morita, K. Murakami, M. Nagamatu, R. Nartallo, P. Nieminen, T. Nishimura, K. Ohtsubo, M. Okamura, S. O'Neale, Y. Oohata, K. Paech, J. Perl, A. Pfeiffer, M. G. Pia, F. Ranjard, A. Rybin, S. Sadilov, E. Di Salvo, G. Santin, T. Sasaki, N. Savvas, Y. Sawada, S. Scherer, S. Sei, V. Sirotenko, D. Smith, N. Starkov, H. Stoecker, J. Sulkimo, M. Takahata, S. Tanaka, E. Tcherniaev, E. Safai Tehrani, M. Tropeano, P. Truscott, H. Uno, L. Urban, P. Urban, M. Verderi, A. Walkden, W. Wander, H. Weber, J. P. Wellisch, T. Wenaus, D. C. Williams, D. Wright, T. Yamada, H. Yoshida, and D. Zschiesche, \href{https://doi.org/10.1016/S0168-9002(03)01368-8}{Nucl. Instrum. Methods Phys. Res. A \textbf{506}, 250 (2003).}

\bibitem{Allison_NIMA2016}
J. Allison, K. Amako, J. Apostolakis, P. Arce, M. Asai, T. Aso, E. Bagli, A. Bagulya, S. Banerjee, G. Barrand, B.R. Beck, A.G. Bogdanov, D. Brandt, J.M.C. Brown, H. Burkhardt, Ph. Canal, D. Cano-Ott, S. Chauvie, K. Cho, G.A.P. Cirrone, G. Cooperman, M.A. Cort\'{e}s-Giraldo, G. Cosmo, G. Cuttone, G. Depaola, L. Desorgher, X. Dong, A. Dotti, V.D. Elvira, G. Folger, Z. Francis, A. Galoyan, L. Garnier, M. Gayer, K.L. Genser, V.M. Grichine, S. Guatelli, P. Gu\`{e}ye, P. Gumplinger, A.S. Howard, I. H\v{r}ivn\'{a}\v{c}ov\'{a}, S. Hwang, S. Incerti, A. Ivanchenko, V.N. Ivanchenko, F.W. Jones, S.Y. Jun, P. Kaitaniemi, N. Karakatsanis, M. Karamitros, M. Kelsey, A. Kimura, T. Koi, H. Kurashige, A. Lechner, S.B. Lee, F. Longo, M. Maire, D. Mancusi, A. Mantero, E. Mendoza, B. Morgan, K. Murakami, T. Nikitina, L. Pandola, P. Paprocki, J. Perl, I. Petrovi\'{c}, M.G. Pia, W. Pokorski, J.M. Quesada, M. Raine, M.A. Reis, A. Ribon, A. Risti\'{c} Fira, F. Romano, G. Russo, G. Santin, T. Sasaki, D. Sawkey, J.I. Shin, I.I. Strakovsky, A. Taborda, S. Tanaka, B. Tom\'{e}, T. Toshito, H.N. Tran, P.R. Truscott, L. Urban, V. Uzhinsky, J.M. Verbeke, M. Verderi, B.L. Wendt, H. Wenzel, D.H. Wright, D.M. Wright, T. Yamashita, J. Yarba, H. Yoshida, \href{https://doi.org/10.1016/j.nima.2016.06.125}{Nucl. Instrum. Methods Phys. Res. A \textbf{835}, 186 (2016).}

\bibitem{Wang_CPC2021}
M. Wang, W. J. Huang, F. G. Kondev, G. Audi, and S. Naimi, \href{https://doi.org/10.1088/1674-1137/abddaf}{Chin. Phys. C \textbf{45}, 030003 (2021).}

\bibitem{Glassman_PRC2019}
B. E. Glassman, D. P\'{e}rez-Loureiro, C. Wrede, J. Allen, D. W. Bardayan, M. B. Bennett, K. A. Chipps, M. Febbraro, M. Friedman, C. Fry, M. R. Hall, O. Hall, S. N. Liddick, P. O'Malley, W. -J. Ong, S. D. Pain, S. B. Schwartz, P. Shidling, H. Sims, L. J. Sun, P. Thompson, and H. Zhang, \href{https://doi.org/10.1103/PhysRevC.99.065801}{Phys. Rev. C \textbf{99}, 065801 (2019).}

\bibitem{Sun_PRC2021}
L. J. Sun, M. Friedman, T. Budner, D. P\'{e}rez-Loureiro, E. Pollacco, C. Wrede, B. A. Brown, M. Cortesi, C. Fry, B. E. Glassman, J. Heideman, M. Janasik, A. Kruskie, A. Magilligan, M. Roosa, J. Stomps, J. Surbrook, and P. Tiwari, \href{https://doi.org/10.1103/PhysRevC.103.014322}{Phys. Rev. C \textbf{103}, 014322 (2021).}

\bibitem{Chen_NDS2017}
J. Chen, \href{https://doi.org/10.1016/j.nds.2017.02.001}{Nucl. Data Sheets \textbf{140}, 1 (2017).}

\bibitem{Zyla_PTEP2020}
P.A. Zyla \textit{et al}. (Particle Data Group), \href{https://doi.org/10.1093/ptep/ptaa104}{Prog. Theor. Exp. Phys. \textbf{2020}, 083C01 (2020).}

\bibitem{Moss_NPA1970}
C.E. Moss, \href{https://doi.org/10.1016/0375-9474(70)90434-3}{Nucl. Phys. A \textbf{145}, 423 (1970).}

\bibitem{Doherty_PRL2012}
D. T. Doherty, G. Lotay, P. J. Woods, D. Seweryniak, M. P. Carpenter, C. J. Chiara, H. M. David, R. V. F. Janssens, L. Trache, and S. Zhu, \href{https://doi.org/10.1103/PhysRevLett.108.262502}{Phys. Rev. C \textbf{108}, 262502 (2012).}

\bibitem{Doherty_PRC2014}
D. T. Doherty, P. J. Woods, G. Lotay, D. Seweryniak, M. P. Carpenter, C. J. Chiara, H. M. David, R. V. F. Janssens, L. Trache, and S. Zhu, \href{https://doi.org/10.1103/PhysRevC.89.045804}{Phys. Rev. C \textbf{89}, 045804 (2014).}

\bibitem{Kankainen_PLB2017}
A. Kankainen, P.J. Woods, H. Schatz, T. Poxon-Pearson, D.T. Doherty, V. Bader, T. Baugher, D. Bazin, B.A. Brown, J. Browne, A. Estrade, A. Gade, J. Jos\'{e}, A. Kontos, C. Langer, G. Lotay, Z. Meisel, F. Montes, S. Noji, F. Nunes, G. Perdikakis, J. Pereira, F. Recchia, T. Redpath, R. Stroberg, M. Scott, D. Seweryniak, J. Stevens, D. Weisshaar, K. Wimmer, R. Zegers, \href{https://doi.org/10.1016/j.physletb.2017.01.084}{Phys. Lett. B \textbf{769}, 549 (2017).}

\bibitem{Jenkins_PRC2005}
D. G. Jenkins, C. J. Lister, M. P. Carpenter, P. Chowdhury, N. J. Hammond, R. V. F. Janssens, T. L. Khoo, T. Lauritsen, D. Seweryniak, T. Davinson, P. J. Woods, A. Jokinen, and H. Penttila, \href{https://doi.org/10.1103/PhysRevC.72.031303}{Phys. Rev. C \textbf{72}, 031303(R) (2005).}

\bibitem{Jenkins_PRC2006}
D. G. Jenkins, A. Meadowcroft, C. J. Lister, M. P. Carpenter, P. Chowdhury, N. J. Hammond, R. V. F. Janssens, T. L. Khoo, T. Lauritsen, D. Seweryniak, T. Davinson, P. J. Woods, A. Jokinen, H. Penttila, G. Mart\`{i}nez-Pinedo, and J. Jos\'{e}, \href{https://doi.org/10.1103/PhysRevC.73.065802}{Phys. Rev. C \textbf{89}, 045804 (2014).}

\bibitem{Testov_PRC2021}
D. A. Testov, A. Boso, S. M. Lenzi, F. Nowacki, F. Recchia, G. de Angelis, D. Bazzacco, G. Colucci, M. Cottini, F. Galtarossa, A. Goasduff, A. Gozzelino, K. Hady\'{n}ska-Klek, G. Jaworski, P. R. John, S. Lunardi, R. Menegazzo, D. Mengoni, A. Mentana, V. Modamio, A. Nannini, D. R. Napoli, M. Palacz, M. Rocchini, M. Siciliano, and J. J. Valiente-Dob\'{o}n, \href{https://doi.org/ 10.1103/PhysRevC.104.024309}{Phys. Rev. C \textbf{104}, 024309 (2021).}

\bibitem{Setoodehnia_PRC2020}
K. Setoodehnia, A. A. Chen, J. Chen, J. A. Clark, C. M. Deibel, J. Hendriks, D. Kahl, W. N. Lennard, P. D. Parker, D. Seiler, and C. Wrede, \href{https://doi.org/10.1103/PhysRevC.102.045806}{Phys. Rev. C \textbf{102}, 045806 (2020).}

\bibitem{ICRU73_2005}
ICRU Report 73, \textit{Stopping of Ions Heavier than Helium}, \href{https://doi.org/10.1093/jicru_ndi001}{International Commission on Radiation Units and Measurements, Oxford University Press, Oxford, 2005.}

\bibitem{Rodgers_NIMA2021}
Cade R. Rodgers and Christian Iliadis, \href{https://doi.org/10.1016/j.nima.2021.165172}{Nucl. Instrum. Methods Phys. Res. A \textbf{998}, 165172 (2021).}

\bibitem{Tipping_NC1999}
Michael E. Tipping and Christopher M. Bishop, \href{https://doi.org/10.1162/089976699300016728}{Neural Comput. \textbf{11}, 443 (1999).}

\bibitem{Rasmussen_2006}
C. E. Rasmussen and C. K. I. Williams, \textit{Gaussian Processes for Machine Learning} (Cambridge, MA: The MIT Press, 2006).

\bibitem{MADAI}
\href{http://madai.us/}{MADAI: Modeling and Data Analysis Initiative}.

\bibitem{JCGM_2008}
Evaluation of measurement data - Supplement 1 to the \textit{Guide to the expression of uncertainty in measurement - Propagation of distributions using a Monte Carlo method}. \href{https://www.bipm.org/documents/20126/2071204/JCGM_101_2008_E.pdf/325dcaad-c15a-407c-1105-8b7f322d651c}{Joint Committee for Guides in Metrology,
101, 2008.}

\bibitem{Brown_NDS2014}
B.A. Brown and W.D.M. Rae, \href{https://doi.org/10.1016/j.nds.2014.07.022}{Nucl. Data Sheets \textbf{120}, 115 (2014).}

\bibitem{Brown_PRC2006}
B. Alex Brown and W. A. Richter, \href{https://doi.org/10.1103/PhysRevC.74.034315}{Phys. Rev. C \textbf{74}, 034315 (2006).}

\bibitem{Magilligan_PRC2020}
A. Magilligan and B. A. Brown, \href{https://doi.org/10.1103/PhysRevC.101.064312}{Phys. Rev. C \textbf{101}, 064312 (2020).}

\bibitem{Richter_PRC2008}
W. A. Richter, S. Mkhize, and B. A. Brown, \href{https://doi.org/10.1103/PhysRevC.78.064302}{Phys. Rev. C \textbf{78}, 064302 (2008).}

\bibitem{Ouellet_NDS2013}
Christian Ouellet and Balraj Singh, \href{https://doi.org/10.1016/j.nds.2013.03.001}{Nucl. Data Sheets \textbf{114}, 209 (2013).}

\bibitem{Brown_PRC2014}
B. Alex Brown, W. A. Richter, and C. Wrede, \href{https://doi.org/10.1103/PhysRevC.89.062801}{Phys. Rev. C \textbf{89}, 062801(R) (2014).}

\bibitem{Basunia_NDS2010}
M. Shamsuzzoha Basunia, \href{https://doi.org/10.1016/j.nds.2010.09.001}{Nucl. Data Sheets \textbf{111}, 2331 (2010).}

\end{thebibliography}
\end{document}